\documentclass[12pt]{iopart}
\begin{document}

\letter{Griffiths inequalities for the Gaussian spin glass}

\author{Satoshi Morita\dag, Hidetoshi Nishimori\dag 
\ and Pierluigi Contucci\ddag}

\address{\dag\ Department of Physics, Tokyo Institute of Technology,
Oh-okayama, Meguro-ku, Tokyo 152-8551, Japan}

\address{\ddag\ Dipartimento di Matematica, Universit\`a di Bologna,
40127 Bologna, Italy}

\begin{abstract}
 The Griffiths inequalities for Ising spin-glass models with Gaussian
 randomness of non-vanishing mean are proved using properties of the
 Gaussian distribution and gauge symmetry of the system. These
 inequalities imply that correlation functions are non-negative and
 monotonic along the Nishimori line in the phase diagram. From this
 result, the existence of thermodynamic limit for correlation
 functions and pressure is proved under free and fixed boundary
 conditions. Relations between the location of multicritical points are
 also derived for different lattices.
\end{abstract}

% Uncomment for PACS numbers title message


%\maketitle

\section{Introduction}
Two Griffiths inequalities provide a significant insight about the
phase transitions in ferromagnetic Ising models \cite{domb}. One of
their formulations states that if the sets of Ising spins
$S_A=\prod_{i\in A}S_i$ are coupled by an energy $- J_A S_A$ (for some
positive $J_A$), the free energy $F$ and all the correlations are
monotonic functions of the strength of any interactions $J$'s,
namely:
\begin{equation}
- \frac{\rmd}{\rmd J_A} F \geq 0 \; ,
 \label{GF1}
\end{equation}
\begin{equation}
\frac{\rmd}{\rmd J_B} \langle S_A \rangle \geq 0 \; .
 \label{GF2}
\end{equation}
These inequalities can be used in ferromagnetic Ising models to prove
that the free energy and correlation functions have the thermodynamic
limit under several boundary conditions and to demonstrate the existence
of phase transitions for various lattices. The proof of equations
(\ref{GF1}) and (\ref{GF2}) \cite{griff}
assumes that all the interactions among spins are ferromagnetic, a
condition that clearly fails in spin glass models which have both
ferromagnetic and antiferromagnetic interactions.

Attempts to extend
even only partially those or similar inequalities to the spin glasses
have been unsuccessful until very recently after the technique of
integration by parts has been powerfully exploited in the mathematically
rigorous approaches: in \cite{gt,cg1} the main results are based on
correlation inequalities that represents the equivalent of the {\it
first} Griffiths inequality (\ref{GF1}) for the Sherrington-Kirkpatrick model and
the Edwards-Anderson model respectively.  The monotonicity properties in
\cite{gt,cg1} are proved not with respect to the strength of the
interaction but with respect to the {\it variance} of the random
interaction.

In this paper we show that a Gaussian Ising spin-glass model does
fulfil the {\it first} and the {\it second} Griffiths inequalities (\ref{GF1})
and (\ref{GF2}) with
respect to the {\it mean} of the distribution: pressure and correlations
are monotonic functions with respect to the mean. The results are proved on the
Nishimori line (NL), a restricted space of the phase diagram in which
several exact results follow from the gauge symmetries of the system
\cite{nishi_ptp,nishi_ox}.

We apply the resulting inequalities to prove the existence of the
thermodynamic limit for the pressure and correlation functions under
free and fixed boundary conditions. Moreover, we derive inequalities on
the location of multicritical points.

In the next section, we present our results and their proofs. The
applications of the inequalities are discussed in the last
section. Some of the details of calculations are described in the
appendix.

\section{Inequalities}

Let us consider a finite box $\Omega$, subset of a regular lattice 
${\cal L}$. To each point $i\in \Omega$ we associate an Ising spin
$(S_i =
\pm 1)$ and denote their products by
\begin{equation}
 S_A = \prod_{i \in A} S_i \; .
\end{equation}
We consider the spin glass model defined by the random potential
\begin{equation}
 U_\Omega = \sum_{A \subset \Omega} \beta_A J_A S_A \; ,
\end{equation}
where $\beta_A\geq 0$ is the inverse of local temperature of subset $A$
and $J_A$ is a quenched random variable which follows the Gaussian
distribution with positive mean $[J_A]=J_{A0}$ and variance
$[(J_A-J_{A0})^2]=\sigma^2_A$.
The {\it pressure} function is defined as
\begin{equation}
 P = [\; \log{\sum_S}\rme^{U_\Omega}\; ]\;,
\label{pressure}
\end{equation} 
from which all physical quantities can be derived. 

The NL is defined in terms of the parameters $x_A\ge 0$ (for all the $A$) as
\cite{nishi_ptp,nishi_ox}
\begin{equation}
 \beta_A = \frac{x_A}{\sigma_A}\;,\quad J_{A0}= \sigma_A x_A \; .
 \label{eq:nl_condition}
\end{equation}
Our results are the following inequalities:
\begin{equation}
\frac{\rmd P}{\rmd x_B} = x_B[\langle S_B +1\rangle] 
\geq 0
 \label{g1}
\end{equation}
\begin{equation}
 \frac{\rmd}{\rmd x_B}[\langle S_C \rangle] = 2x_B [ (\langle S_B S_C \rangle
  - \langle S_B \rangle \langle S_C \rangle)^2] \geq 0 \; .\label{g2}
\end{equation}
Both inequalities hold
for arbitrary subsets $B,C$ as long as the parameters satisfy NL
condition (\ref{eq:nl_condition}).

The first and second inequalities are proved using some
properties of the Gaussian distribution and the gauge theory. 
To prove the first inequality (\ref{g1}), we observe that on the NL
the formula of the {\it total derivative} gives
\begin{equation}
 \frac{\rmd P}{\rmd x_B}
  =  \frac{\partial P}{\partial \beta_B} \frac{\rmd\beta_B}{\rmd x_B}
  + \frac{\partial P}{\partial J_{B0}}\frac{\rmd J_{B0}}{\rmd x_B} =
\left.\frac{1}{\sigma_B}\frac{\partial P}{\partial
\beta_B}\right|_{\rm NL}
  + \left.\sigma_B\frac{\partial P}{\partial J_{B0}}\right|_{\rm NL} .
  \label{td}
\end{equation}
As shown in the appendix, we can derive the following identities from the
properties of the Gaussian distribution for any operator $O$:
\begin{eqnarray}
 [J_B O]=J_{B0}[O] +\sigma_B^2 \left[\frac{\partial O}{\partial J_B}\right] 
 \label{eq:gausseq1} \\
 \frac{\partial}{\partial J_{B0}} [O]
  = \left[ \frac{\partial O}{\partial J_B} \right] . 
 \label{eq:gausseq2}
\end{eqnarray}
Note that both equations (\ref{eq:gausseq1}) and (\ref{eq:gausseq2}) are
valid on and away from the NL.  The identity (\ref{eq:gausseq2}) yields
\begin{equation}
\frac{\partial P}{\partial J_{B0}} = \beta_B[\langle S_B\rangle] \; .
\label{dpdJ0}
\end{equation}
and a direct computation provides
\begin{equation}
\frac{\partial P}{\partial \beta_B} =  [J_B\langle S_B\rangle] \; .
\end{equation}
Using the gauge theory we have on the NL
\cite{nishi_ptp,nishi_ox}
\begin{equation}
 [\langle J_B S_B \rangle] = J_{B0} = \sigma_B x_B \; . 
\label{eq:energy}
\end{equation}
From equations (\ref{td}), (\ref{dpdJ0})-(\ref{eq:energy}),
we immediately find equation (\ref{g1}).

It is also possible to obtain a related inequality, an
explicit bound for the correlation function, as
\begin{equation}
 [ \langle S_B \rangle] \geq \frac{x_B^2}{1+x_B^2}  \quad (B\in\{A\}) \; ,
  \qquad [ \langle S_B \rangle]\geq 0 \quad (B\not\in\{A\}) \; .
\label{lb}
\end{equation}
For this purpose we observe that, when $B\in\{A\}$,
\begin{equation}
 [\langle J_B S_B \rangle] = \int \prod_{A\subset\Omega} \rmd J_A
  \left(\sqrt{P_A(J_A)} J_B\right)
  \left(\sqrt{P_A(J_A)}\langle S_B \rangle \right).
\end{equation}
Let us square both sides of the above equation and apply the
Cauchy-Schwarz inequality to obtain
\begin{equation}
 [\langle J_B S_B\rangle]^2 \leq (\sigma_B^2 +J_{B0}^2)
  [\langle S_B\rangle^2] \; . 
\label{eq:cauchy}
\end{equation}
Since the gauge theory yields an identity \cite{nishi_ptp,nishi_ox}
\begin{equation}
 [\langle S_B\rangle] =[\langle S_B\rangle^2]
\label{eq:m=q}
\end{equation}
on the NL under certain boundary conditions (free, periodic or fixed
(all spins up)), we obtain equation (\ref{lb}) from equations (\ref{eq:energy})
and (\ref{eq:cauchy}).  When $B\not\in\{A\}$, the inequality (\ref{lb})
is a direct consequence of the identity (\ref{eq:m=q}).  We note that
the Gaussian distribution is not essential for the proof of the above
inequality (\ref{lb}).

The proof of the second inequality (\ref{g2}) is similarly carried out.
Derivative by the parameter $x_B$ is expressed as in equation
(\ref{td}) by
\begin{equation}
 \frac{\rmd}{\rmd x_B} [\langle S_C \rangle]
  = \left.\frac{1}{\sigma_B}\frac{\partial}{\partial \beta_B}
  [\langle S_C \rangle]\right|_{\rm NL} 
  + \left.\sigma_B\frac{\partial}{\partial J_{B0}}
  [\langle S_C \rangle]\right|_{\rm NL} .
\label{eq:div_x}
\end{equation}
Derivative by $\beta_B$ is easily calculated as
\begin{equation}
 \frac{\partial}{\partial \beta_B}[\langle S_C \rangle]
  = [J_B \langle S_B S_C \rangle]
  - [J_B \langle S_B\rangle\langle  S_C \rangle] \; . 
\label{eq:div_beta}
\end{equation}
Substitution of equation (\ref{eq:gausseq1}) into equation (\ref{eq:div_beta}) yields
\begin{equation}
 \eqalign{
 \frac{1}{\sigma_B}\frac{\partial}{\partial \beta_B}
  [\langle S_C \rangle] &= \frac{J_{B0}}{\sigma_B} [\langle S_B S_C \rangle
  - \langle S_B \rangle \langle S_C \rangle ]
 \label{eq:div_beta2} \\
  &\quad -2 \sigma_B \beta_B [\langle S_B \rangle \langle S_B S_C \rangle
  - \langle S_B \rangle^2 \langle S_C \rangle ] \; ,
 }
\end{equation}
and the identity (\ref{eq:gausseq2}) gives
\begin{equation}
 \sigma_B \frac{\partial}{\partial J_{B0}}[\langle S_C \rangle]
  = \sigma_B \beta_B [\langle S_B S_C \rangle
  - \langle S_B \rangle \langle S_C \rangle] \; .
\label{eq:div_J0}
\end{equation}
From equation (\ref{eq:div_x}) we obtain
\begin{equation}
 \frac{\rmd}{\rmd x_B}[\langle S_C \rangle]
  =2x_B [\langle S_B S_C \rangle
  - \langle S_B \rangle \langle S_C \rangle
  - \langle S_B \rangle \langle S_B S_C \rangle
  + \langle S_B \rangle^2 \langle S_C \rangle ] \; .
\label{eq:div_x_S}
\end{equation}
The gauge theory yields the following identities on the NL under the
same boundary conditions as the identity (\ref{eq:m=q})
\cite{nishi_ptp,nishi_ox}:
\begin{equation}
 \eqalign{
    &[ \langle S_B S_C \rangle] = [\langle S_B S_C \rangle^2] \\
  &[ \langle S_B \rangle \langle S_C \rangle]
  = [ \langle S_B \rangle \langle S_B S_C \rangle]
  = [ \langle S_B \rangle \langle S_C \rangle \langle S_B S_C \rangle] \\
  &[ \langle S_B \rangle^2 \langle S_C \rangle]
  =[ \langle S_B \rangle ^2 \langle S_C \rangle^2] \; .
 }
\end{equation}
Therefore, we obtain from equation (\ref{eq:div_x_S})
\begin{equation}
 \frac{\rmd}{\rmd x_B}[\langle S_C \rangle] = 2x_B [(\langle S_B S_C \rangle
  - \langle S_B \rangle \langle S_C \rangle)^2] \geq 0 \; .
\end{equation}
In general, the following relation is proved similarly:
\begin{equation}
 \eqalign{
  &\frac{\rmd}{\rmd x_B}[\langle S_C \rangle^{2k-1}] 
  = \frac{\rmd}{\rmd x_B}[\langle S_C \rangle^{2k}] \\ 
  &=2k(2k-1)x_B [ \langle S_B \rangle^{2k-2}(\langle S_B S_C \rangle
  - \langle S_B \rangle \langle S_C \rangle)^2] \geq 0 \; .
 }
\end{equation}

It is also possible to prove concavity of the pressure
from the derivative of equation (\ref{g1}), since the second
derivative of the pressure is calculated as
\begin{equation}
 \frac{\rmd^2 P}{\rmd x_B \rmd x_C} =
  \cases{x_B \frac{\rmd}{\rmd x_C}[\langle S_B \rangle]
  \; \geq \; 0 & ($B\neq C$)\\
  [\langle S_B \rangle]+1+x_B \frac{\rmd}{\rmd x_B}[\langle S_B \rangle] 
  \; \geq \; 0 & ($B=C$)}
\end{equation}
where we use equations (\ref{g2}) and (\ref{lb}).

\section{Discussions on physical consequences}

The inequality (\ref{lb}) implies that a correlation
function $[\langle S_B \rangle]$ is non-negative and increases toward
unity when the parameter $x_B$ of the corresponding subset $B$ tends to
infinity.  This result is reasonable: for large $x_B$, the interaction
$J_B$ is almost ferromagnetic and the local temperature is nearly zero,
and therefore all the spins in the subset $B$ become parallel to each
other.

An immediate result from the second inequality (\ref{g2}) is that an
arbitrary $n$-point correlation function, including order parameters, is
an increasing function of a parameter for any subset. In particular, a
two-point correlation function increases with $x$ (which is proportional
to the inverse temperature and the centre of distribution) and thus the
correlation length becomes larger as $x$ increases, a natural result.

The two inequalities have profound consequences on the structure and
existence of the thermodynamic limit $\Omega\nearrow {\cal L}$ for both
pressure and correlation functions.  The first inequality (\ref{g1})
tells us that the pressure is monotonically increasing with any
$x_A$. Since on the NL the Boltzmann weight admits the representation
\begin{equation}
\rme^{\sum_{A\in\Omega}x_Aj_AS_A} \; ,
\label{bnl}
\end{equation}
where $j_A=J_A/\sigma_A$ is a Gaussian variable with mean $[j_A]=x_A\ge
0$ and variance $[(j_A-x_A)^2]=1$, the pressure can be expressed as
\begin{equation}
P = [\; \log{\sum_S}\rme^{\sum_{A\in\Omega}x_Aj_AS_A}\;] \; ,
\label{pressure_n}
\end{equation}
and the correlation as 
\begin{equation}
[\langle S_C \rangle] = 
\left[ \frac{\sum_{S}S_C\rme^{\sum_{A\in\Omega}x_Aj_AS_A}}
{\sum_{S}\rme^{\sum_{A\in\Omega}x_Aj_AS_A}} \right] \; ,
\label{co_rep}
\end{equation}
where both $P$ and $[\langle S_C \rangle]$ are functions only of $x$'s. 
From equation (\ref{pressure_n}) we see that the $x_A$ {\it tune}
the interaction of the set of spins $A$.

We can show the existence of the thermodynamic limit for the pressure
per spin under free boundary condition. Take a large cube $\Omega$
(consider for simplicity $d=2$) and cut it in four identical cubes
$\Omega_{i}$, $i=1,...,4$. We can tune {\it off} the interactions among
the cubes simply by setting all the $x$'s among them equal to zero. In
this case the total pressure is
$P_{\Omega_1}+P_{\Omega_2}+P_{\Omega_3}+P_{\Omega_4}$.  Then we can tune
{\it on} the interactions from zero to their original value and the
total pressure becomes $P_\Omega$. The monotonicity property (\ref{g1})
gives
\begin{equation}
P_{\Omega_1}+P_{\Omega_2}+P_{\Omega_3}+P_{\Omega_4} \leq P_\Omega \; ,
\end{equation}
which is the well known sub-additivity property for the pressure and
implies the existence of its density in the thermodynamic limit under
two natural assumptions of invariance by translation of the Gaussian
distributions and the stability boundedness \cite{cg2}.

Moreover the second inequality (\ref{g2}) can be used to prove the
existence of the thermodynamic limit for the correlation functions in
the same way as the first is used to prove the existence of the
pressure. In fact, from equation (\ref{bnl}), we see that when $x_A=0$
there is no interaction among the subset $A$, and to increase $x_A$ from
zero is equivalent to add an interaction to the subset $A$. From the
second inequality (\ref{g2}), addition of an interaction for any subset
increases all correlation functions.

This result can be used for the proof that correlation functions have a
thermodynamic limit under free boundary conditions. Let us consider two
finite sets $\Omega' \subset \Omega$. The subset $\Omega$ is obtained
from $\Omega'$ by adding interactions. Then it is clear from the previous
argument that
\begin{equation}
 [\langle S_B \rangle]_{\Omega'}^{(\rm free)} \leq 
 [\langle S_B \rangle]_{\Omega}^{(\rm free)} .
\end{equation}
This implies that the correlation functions monotonically increase with
the system size.  Since correlation functions are bounded by unity, each
of them tends to its unique limit as the volume of $\Omega$ tends to
infinity. Therefore the thermodynamic limit for correlation functions
exists on the NL.

Monotonicity of the correlation functions with system size can also be
proved under fixed boundary condition.  Fixed boundary conditions ($S_i
= +1$ for all boundary sites $i\in \partial \Omega$) is represented by
applying very strong positive magnetic fields to all sites outside
$\Omega$. This is equivalent to $x_A\rightarrow\infty$ for $A=\{i\}$,
$i\in\partial\Omega$. Since the set $\Omega \supset \Omega'$ is obtained
from $\Omega'$ by reducing magnetic fields for sites $i\in \Omega
\setminus\Omega'$, we have
\begin{equation}
 [\langle S_B \rangle]_{\Omega'}^{(\rm fix)} 
  \geq [\langle S_B \rangle]_{\Omega}^{(\rm fix)}
\end{equation}
from the inequality (\ref{g2}). Because the inequality
(\ref{lb}) yields lower bounds for $[\langle S_B \rangle]$,
there should exist a well-defined limit as $\Omega\nearrow {\cal L}$. The
thermodynamic limit of $[\langle S_B \rangle]^{(\rm fix)}$ may not
necessarily be equal to that of $[\langle S_B \rangle]^{(\rm free)}$, but
the former cannot be less than the latter.

The second inequality can also be used for the argument about the
location of the multicritical points, which are believed to lie on the
NL \cite{nishi_ptp,nishi_ox}, for various lattices. For example, let us
consider three two-dimensional lattices, the triangular (TR), square
(SQ) and hexagonal (HEX) lattices. The triangular lattice is obtained
from the square lattice with addition of bonds, and the square lattice
is obtained from the hexagonal. Thus, the magnetizations of three
lattices satisfy the following relation:
\begin{equation}
 [\langle S_i \rangle]_{\rm HEX} 
  \leq [\langle S_i \rangle]_{\rm SQ} 
  \leq [\langle S_i \rangle]_{\rm TR} \; .
\end{equation}
Since the multicritical temperature is defined as
\begin{equation}
 T_c = \sup \{T; [\langle S_i \rangle]>0\} \; ,
\end{equation}
we obtain
\begin{equation}
 T_c^{\rm HEX} \leq T_c^{\rm SQ} \leq  T_c^{\rm TR} \; .
\end{equation}
Similarly, the multicritical temperature of the simple cubic lattice is
higher then that of the square lattice because the former lattice is
constructed from the latter by adding interactions.

It is an interesting future problem to prove our results for other
models, for example, the $\pm J$ Ising model, and to develop similar
analyses away from the NL.

After submission of the manuscript we learnt that Kitatani
\cite{kitatani} had discussed a related problem.

\ack This work was supported by the Grant-in-Aid for Scientific Research
on Priority Area ``Statistical-Mechanical Approach to Probabilistic
Information Processing'' by the MEXT and by the 21st Century COE Program
``Nanometer-Scale Quantum Physics'' at Tokyo Institute of
Technology. One of us (P.C.) thanks Tokyo Institute of Technology for
the hospitality.

\appendix
\setcounter{section}{1}
\section*{Appendix}

In this appendix, we derive equations (\ref{eq:gausseq1}) and
(\ref{eq:gausseq2}) using some properties of the Gaussian distribution
\begin{equation}
P_A(J_A) = \frac{1}{\sqrt{2\pi\sigma_A^2}}
  \exp\!\left(-\frac{(J_A-J_{A0})^2}{2\sigma_A^2}\right) \; .
\end{equation}
Let us consider the following quantity,
\begin{equation}
 \left[\frac{\partial}{\partial J_B} O(\{J_A\})\right]
  = \int \prod_{A\subset\Omega}\rmd J_A P_A(J_A)
  \frac{\partial}{\partial J_B} O(\{J_A\}) \; ,
\end{equation}
where $O(\{J_A\})$ is a function of interactions $\{J_A\}$. Since
Gaussian distribution $P_A(J_A)$ decays rapidly as
$|J_A|\rightarrow\infty$, we may rewrite the right-hand side using
integration by parts as
\begin{eqnarray}
 \left[\frac{\partial}{\partial J_B} O(\{J_A\})\right]
 &= \int\prod_{\stackrel{\scriptstyle A\subset\Omega}{A\neq B}} 
  \rmd J_A P_A(J_A) \rmd J_B P_B(J_B)
  \frac{\partial}{\partial J_B} O(\{J_A\}) \\
 &= \int\prod_{\stackrel{\scriptstyle A\subset\Omega}{A\neq B}} 
  \rmd J_A P_A(J_A) \rmd J_B \biggl( \!-\frac{\partial}{\partial J_B}
  P_B(J_B) \!\biggr) O(\{J_A\}) \; .
\label{div_JB}
\end{eqnarray}
Calculating the derivative of the Gaussian yields
\begin{equation}
 \frac{\partial}{\partial J_B}P_B(J_B)
  = \frac{J_{B0}-J_B}{\sigma_B^2}P_B(J_B) \; .
\end{equation}
This relation shows that
\begin{eqnarray}
 \left[\frac{\partial}{\partial J_B} O(\{J_A\})\right]
 &= \int\prod_{A\subset\Omega} \rmd J_A P_A(J_A)
 \frac{J_B-J_{B0}}{\sigma_B^2} O(\{J_A\}) \\
 &= \frac{1}{\sigma_B^2} \Bigl\{ [J_B  O(\{J_A\})]
 - J_{B0}[O(\{J_A\})]\Bigr\}
\end{eqnarray}
which is equation (\ref{eq:gausseq1}).

For the proof of (\ref{eq:gausseq2}), we use the property that the Gaussian
distribution $P_B(J_B)$ is a function of $J_B-J_{B0}$,
\begin{equation}
 \frac{\partial}{\partial J_B} P_B(J_B)
  = -\frac{\partial}{\partial J_{B0}} P_B(J_B) \; .
\label{eq:change_div}
\end{equation}
Substitution of (\ref{eq:change_div}) into (\ref{div_JB}) yields
\begin{equation}
\eqalign{
 \left[\frac{\partial}{\partial J_B} O(\{J_A\})\right]
 &= \int\prod_{\stackrel{\scriptstyle A\subset\Omega}{A\neq B}}
  \rmd J_A P_A(J_A) \rmd J_B \left(\frac{\partial}{\partial J_{B0}}
  P_B(J_B) \right) O(\{J_A\}) \\
 &= \frac{\partial}{\partial J_{B0}}
  \int\prod_{A\subset\Omega} \rmd J_A P_A(J_A) O(\{J_A\}) \\
 &= \frac{\partial}{\partial J_{B0}} [ O(\{J_A\})] \; .
}
\end{equation}

\end{document}